\documentstyle[emulateapj,psfig,flushrt,tighten]{article}

\def\kms{\,{\rm km\,s^{-1}}}

\def\msun{\,{\rm M_\odot}}

\def\etal{{et al.\ }}

\newcommand\beq{\begin{equation}}
\newcommand\eeq{\end{equation}}
\newcommand{\ba}{\begin{eqnarray}}
\newcommand{\ea}{\end{eqnarray}}  
\def\spose#1{\hbox to 0pt{#1\hss}}
\def\lta{\mathrel{\spose{\lower 3pt\hbox{$\mathchar"218$}}
     \raise 2.0pt\hbox{$\mathchar"13C$}}}
\def\gta{\mathrel{\spose{\lower 3pt\hbox{$\mathchar"218$}}
     \raise 2.0pt\hbox{$\mathchar"13E$}}}

\righthead{Massive black hole binaries}
\submitted{ApJ, accepted}
\makeatletter
\newenvironment{tablehere}
  {\def\@captype{table}}
  {}
\newenvironment{figurehere}
  {\def\@captype{figure}}
  {}
\makeatother
\begin{document}

\title{Interaction of massive black hole binaries with their stellar environment: I. Ejection of hypervelocity stars}

\author{Alberto Sesana\altaffilmark{1}, Francesco Haardt\altaffilmark{1}, \&
Piero Madau\altaffilmark{2}}

\altaffiltext{1}{Dipartimento di Scienze, Universit\'a dell'Insubria, via 
Valleggio  11, 22100 Como, Italy.}
\altaffiltext{2}{Department of Astronomy \& Astrophysics, University of 
California, 1156 High Street, Santa Cruz, CA 95064.}

\begin{abstract}
We use full three-body scattering experiments to study the ejection of hypervelocity
stars (HVSs) by massive black hole binaries (MBHBs) at the center of galaxies.
Ambient stars drawn from a Maxwellian distribution unbound to the binary 
are expelled by the gravitational slingshot. Accurate measurements of thermally 
averaged hardening, mass ejection, and eccentricity growth rates ($H, J,$ and 
$K$) for MBHBs in a fixed stellar background are obtained by numerical 
orbit integration from initial conditions 
determined by Monte Carlo techniques. Three-body interactions create a 
subpopulation of HVSs on nearly radial orbits, with a spatial distribution 
that is initially highly flattened in the inspiral plane of the MBHB, but 
becomes more isotropic with decreasing binary separation. The degree of 
anisotropy is smaller for unequal mass binaries and larger for stars with 
higher kick velocities. Eccentric MBHBs produce a more prominent tail of 
high-velocity stars and break planar symmetry, ejecting HVSs along a 
broad jet perpendicular to the semimajor axis. The jet two-sidedness 
decreases with increasing binary mass ratio, while the jet opening- angle
increases with decreasing kick velocity and orbital separation.     
The detection of a numerous population of HVSs in the halo of the Milky Way 
by the next generation of large astrometric surveys like GAIA may provide 
a unique signature of the history, nature, and environment of the MBH 
at the Galactic center.
\end{abstract}


\section{Introduction}

Massive black holes (MBHs) are a ubiquitous component of nearby galaxy nuclei
(e.g. Magorrian \etal 1998), and galaxies experience multiple hierarchical mergers 
during their lifetime. Following the merger of two halo$+$MBH systems of comparable 
mass (``major mergers''), dynamical friction is known to effectively drag in the 
satellite halo (and its 
MBH) toward the center of the more massive progenitor: this will lead to the 
formation of a bound MBH binary (MBHB) in the violently relaxed core of the newly 
merged stellar system (Begelman, Blandford, \& Rees 1980). Even in the case of 
unequal-mass mergers, gas cooling appears to facilitate the pairing process by 
increasing the resilience of the companion galaxy to tidal disruption (Kazantzidis
\etal 2005). It is expected then that many galaxies will host wide MBHBs during
cosmic history (e.g. Volonteri, Haardt, \& Madau 2003). As the binary 
separation decays, the effectiveness of dynamical friction slowly declines because 
distant stars perturb the binary's center of mass but not its semi-major axis. 
The bound pair then loses orbital energy by capturing stars passing in its 
immediate vicinity and ejecting them at much higher velocities (gravitational slingshot).

It was first pointed out by Hills (1988) that the tidal breakup of binary stars 
by a MBH at the Galactic center may eject one member of the binary with velocities 
$\sim 1000\,\kms$. Such `hypervelocity stars'' (HVSs) 
are also produced by three-body interactions between ambient stars with low 
angular momentum orbits and a ``hard'' MBHB, i.e. a binary whose binding energy 
per unit mass exceeds the star specific kinetic energy.
Assuming SgrA$^*$ to be one component of a MBHB, Yu \& Tremaine (2003)
estimated the number of HVSs expected within the solar radius to be $\sim 10^3$.
Brown \etal (2005) reported the first discovery of a HVS (with a Galactic rest-frame
velocity in excess of $700\,\kms$) in the Galactic halo. This and five more HVSs,
recently discovered by Hirsch \etal (2005) and Brown \etal (2006a, 2006b), are 
all consistent with a Galactic center origin, while an ejection from the LMC 
is more plausible for the seventh HVS currently known (Edelmann \etal 2005).     
Holley-Bockelmann \etal (2005) have proposed that the anomalously fast intracluster
planetary nebulae identified in the Virgo Cluster by Arnaboldi \etal (2004) 
may also be associated with close three-body interactions with a MBHB at the
center of M87. Unbound HVSs travel with velocities so extreme that dynamical ejection from 
a relativistic potential is the most plausible origin, and are becoming increasingly 
recognized as an important tool for understanding the history, nature, and
environment of nuclear MBHs. 

Stars expelled by a MBHB are expected to form a subpopulation with very distinct 
kinematics (e.g. Quinlan 1996, hereafter Q96) as well as spatial structure (Zier
\& Biermann 2001). The phase-space distribution of HVSs ejected by an intermediate-mass 
black hole (IMBH) 
inspiralling into SgrA$^*$ has been recently studied analytically by Levin (2005).   
Stars bound to SgrA$^*$ and drawn from an isotropic cusp are ejected in a burst lasting 
a few dynamical friction timescales: most stars are expelled isotropically if the
inspiral is circular, or in a broad ``jet'' aligned with the IMBH velocity at 
pericenter if the inspiral is eccentric (Levin 2005). Yet most HVSs are produced
during the phases of the inspiral that cannot be modeled analytically.     
This is the first paper in a series aimed at a detailed numerical study of the 
interaction of MBHBs with their dense stellar environment. We use here 
{\it full three-body scattering experiments} of the ejection of background stars 
by MBHBs at the center of galaxies to address the kinematic properties of HVSs.
Ambient stars drawn from a Maxwellian distribution unbound to the binary are expelled 
by the gravitational slingshot. Reaction rates are obtained by numerical orbit 
integration from initial conditions determined by Monte Carlo techniques. 
The plan of the paper is as follows. In \S~2 we describe our suite of three-body 
scattering experiments. 
In \S~3 we present accurate measurements of the binary hardening, mass ejection, 
and eccentricity growth rates for MBHBs embedded in a fixed 
stellar background, reproducing Q96's classical results.  
In \S~4 we discuss the detailed kinematic properties of the ejected 
subpopulation. Finally, we present our conclusions in \S~5.

\section{Scattering experiments}

Consider a binary of mass $M=M_1+M_2$ ($M_2\leq M_1$), reduced mass 
$\mu=M_1M_2/M$, and semimajor axis $a$, orbiting in the $(x,y)$ plane in 
a background of stars of mass $m_*$.
In the case of a light intruder with $m_* \ll M_2$, 
the problem is greatly simplified by setting the center of mass of
the binary {\it at rest} at the origin of the coordinate system. 
It is then convenient to define an approximated  
dimensionless energy change $C$ and angular momentum change $B$ in a single 
binary-star interaction as (Hills 1983)  
\begin{equation}\label{c}
C=\frac{M}{2m_*}\frac{\Delta E}{E}=\frac{a\Delta E_*}{G\mu},
\end{equation}
and 
\begin{equation}\label{b}
B=-\frac{M}{m_*}\frac{\Delta L_z}{L_z}=\frac{M}{\mu}\frac{\Delta L_{z*}}{L_z}. 
\end{equation}
Here $\Delta E/E$ is the fractional increase (decrease if negative) 
in the orbital specific binding energy $E=-GM/(2a)$, $\Delta L_z/L_z$ 
is the fractional 
change in orbital specific angular momentum $L_z=\sqrt{GMa(1-e^2)}$, while
$\Delta E_*$ and $\Delta L_{z*}$ are the corresponding 
changes for the interacting star. Conservation of total energy 
and angular momentum lead to the following 
expression for the change in orbital eccentricity $e$,  
\begin{equation}\label{deltae}
\Delta e=\frac{(1-e^2)}{2e}\frac{2m_*}{M}(B-C).
\end{equation}
The quantities $B$ and $C$ are of order unity and are derived by 
three-body scattering experiments that treat the star-binary encounters
one at a time (Hut \& Bahcall 1983; Q96). For each encounter 
one solves nine coupled, second-order, differential equations,
\begin{equation}\label{eq:3body}
{\ddot{\bf r}_i}=-G\sum_{i\neq j}\frac{m_j({\bf r}_i-{\bf r}_j)}{\mid {\bf r}_i
- {\bf r}_j\mid^3},
\end{equation}
supplied by 18 initial conditions ${\bf r}_i(t=0)$, $\dot{{\bf r}_i}
(t=0)$. 
The incoming star is moved from $r=\infty$ to $r_i=[10^{10}\mu/M]^{1/4}a$
on a Keplerian orbit about a point mass $M$. At $r_i$ the force induced 
by the quadrupole moment of the binary is 10 orders of magnitude 
smaller than the total force acting on the star at a distance $a$, and
numerical integration starts. 
The initial conditions define a point in a nine-dimensional parameter 
space given by:
\begin{itemize}
\item the mass ratio $q=M_2/M_1$ of the binary;
\item the eccentricity of the binary $e$;
\item the mass of the star $m_*$;
\item the asymptotic initial speed of the incoming field star $v\equiv 
|{\bf v}|$;
\item the impact parameter at infinity $b$ (the distance at which the
star would pass the binary if it fell no attraction);
\item four angles: $\theta$ and $\phi$ describing the initial direction of the 
impact, $\psi$ its initial orientation, and $\Psi$ the initial 
binary phase.
\end{itemize}
A significant star-binary energy exchange (i.e. characterized by a dimensionless 
energy change $C>1$) occurs only for 
$v/V_c<\sqrt{M_2/M}$, where $V_c=\sqrt{GM/a}$ is the binary orbital velocity 
(the relative velocity of the two holes if the binary is circular, see, e.g., 
Saslaw, Valtonen, \& Aarseth 1974; Mikkola \& Valtonen 1992). We sample 
this quantity in the range $3\times10^{-3}\sqrt{M_2/M}<v/V_c<30\sqrt{M_2/M}$ 
with 80 logarithmically equally spaced grid points. 
Note that varying the incoming velocity at a fixed 
$V_c$ is equivalent to varying the binary separation $a$ at a fixed $v$.
When averaged over a Maxwellian distribution, the range of $v/V_c$ 
sampled in our experiments therefore probes the slingshot mechanism 
over about four decades in binary separation.

Each scattering experiment requires five random numbers. The impact parameter 
is related to the star pericenter distance $r_p$ by 
\begin{equation}\label{focus}
b^2=r_p^2\left(1+\frac{2GM}{r_p v^2}\right).
\end{equation}
For each incoming velocity the impact parameter is randomly
sampled according to an equal probability distribution in $b^2$ (equivalent
to a probability weight proportional to $b$), in an
interval corresponding to a range in scaled pericenter distance $r_p/a$ of
$[0, 5]$. In a number of test cases we have extended this range to $[0, 10]$,
finding little differences in the measured hardening rates. 
The velocity angles $\theta$ and $\phi$ are randomly 
generated to reproduce a uniform density distribution over a spherical surface
centered in the origin of coordinate system, while the orientation angle $\psi$  
is chosen from a uniform distribution in the range $[0,2\pi]$. The 
probability distribution of the binary phase $\Psi$ is sampled by 
weighting randomly selected angles according to the time spent by the binary  
in any given phase range. Given a set of initial conditions, we integrate
the system of differential equations (\ref{eq:3body}) 
using the most recent version of the subroutine DOPRI5\footnote{See 
http://www.unige.ch/math/folks/hairer/software.html.}, which is based on 
an explicit Runge-Kutta method of order 4(5) due to Dormand \& Prince (1978). A 
complete description of the integrator can be found in Hairer, Norsett, \& 
Wanner (1993).

As the total energy of the binary-star system is always negative, a 
bound triple system may form temporarily, but will typically dissolve 
within 10 to 100 crossing times (Valtonen \& Aarseth 1977).  
The integration is stopped if one of the following events occurs:
(a) the star leaves the sphere of radius $r_i$ with positive total energy; 
(b) the integration reaches $10^6$ time steps (corresponding to 
between a few hundred and a thousand binary orbital periods); or
(c) the physical integration time exceeds $10^{10}$ yrs.
The integration of the full three-body problem allows us to directly control 
the conservation of total energy and angular momentum. The code adjusts the 
integration stepsize to keep the fractional error per step, $\epsilon$, in 
the position and velocity below a level which was set to $10^{-11}$. This 
allows a total energy conservation accuracy $\Delta E/E \sim 10^{-9}$ in 
a single orbit integration, while, for $m_* /M\simeq 10^{-7}$, 
the star energy is conserved, in a single orbit, at level of one part
in a hundred.
We have also checked that our choice of erasing from the record stars 
that get captured for long times does 
not affect results appreciably. Abandoned integrations involve, for the major 
part, encounters with stars that get captured in very weakly bound orbits 
(energetically ``poor''  events) and make many revolutions before being 
expelled. The fraction of erased stars that are instead captured in 
tightly bound orbits (hence energetically ``rich'') is quite small, 
$\lesssim 10^{-4}$, assuring that the global impact of such encounters 
on the evolution of binary orbital parameters is indeed negligible. 
    
We have performed 24 sets of scattering experiments for binary mass ratios 
$q=1$, 1/3, 1/9, 1/27, 1/81, 1/243, and initial eccentricities $e$=0.01, 0.3, 
0.6, 0.9. After each orbit integration the $x,y,z$ components of {\bf v} 
and {\bf L} are stored. Each run involves $4\times10^6$ stars: we collect 
partial outputs after $5\times10^4$ orbit integrations corresponding 
to a given initial speed $v$, calculate the values of $\langle C \rangle$ 
and $\langle B \rangle$ averaged over orbital angular variables, and 
then evaluate the hardening, eccentricity growth, and mass ejection rates 
as defined in the \S~3. Statistical errors are estimated by evaluating the 
rates from ten different orbit subsets, and then computing standard deviations 
(see \S~3).
\begin{figurehere}
\vspace{0.5cm} 
\centerline{\psfig{figure=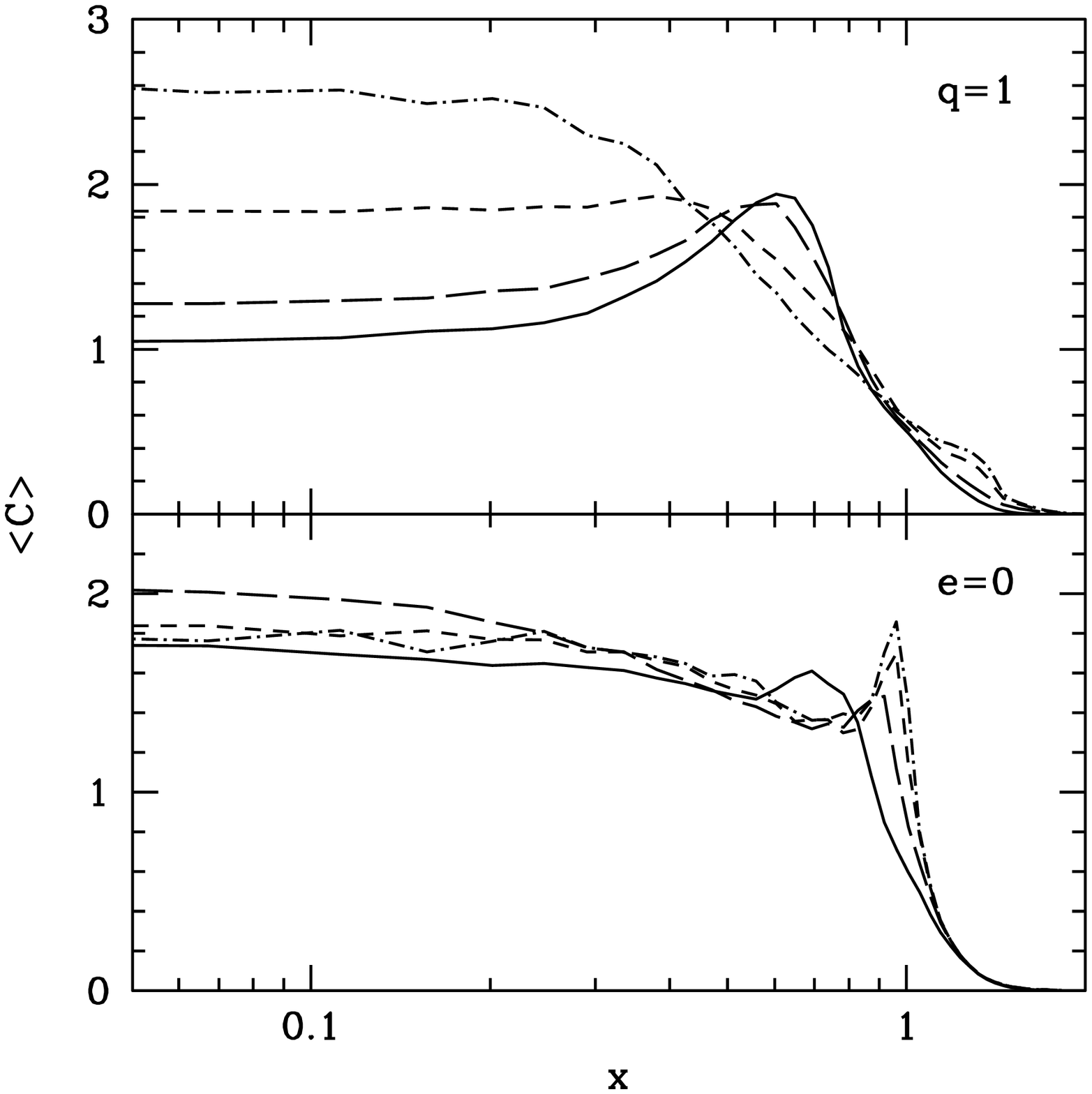,width=2.9in}}
\vspace{-0.0cm}
\caption{\footnotesize {\it Top panel}: mean energy exchange 
$\langle C \rangle$ 
as a function of dimensionless impact parameter $x$ for an equal mass binary 
with eccentricity $e=0$ ({\it solid line}), 0.3 ({\it long-dashed line}), 0.6 
({\it short-dashed line}), and 0.9 ({\it dot-dashed line}). {\it Bottom panel}: same, 
but for a circular binary with mass ratio $q=1/3$ ({\it solid 
line}), 1/9 ({\it long-dashed line}), 1/27 ({\it short-dashed line}), and 1/81 
({\it dot-dashed line}).}
\label{fig:C}
\vspace{+0.5cm}
\end{figurehere}
\begin{figurehere}
\vspace{0.5cm} 
\centerline{\psfig{figure=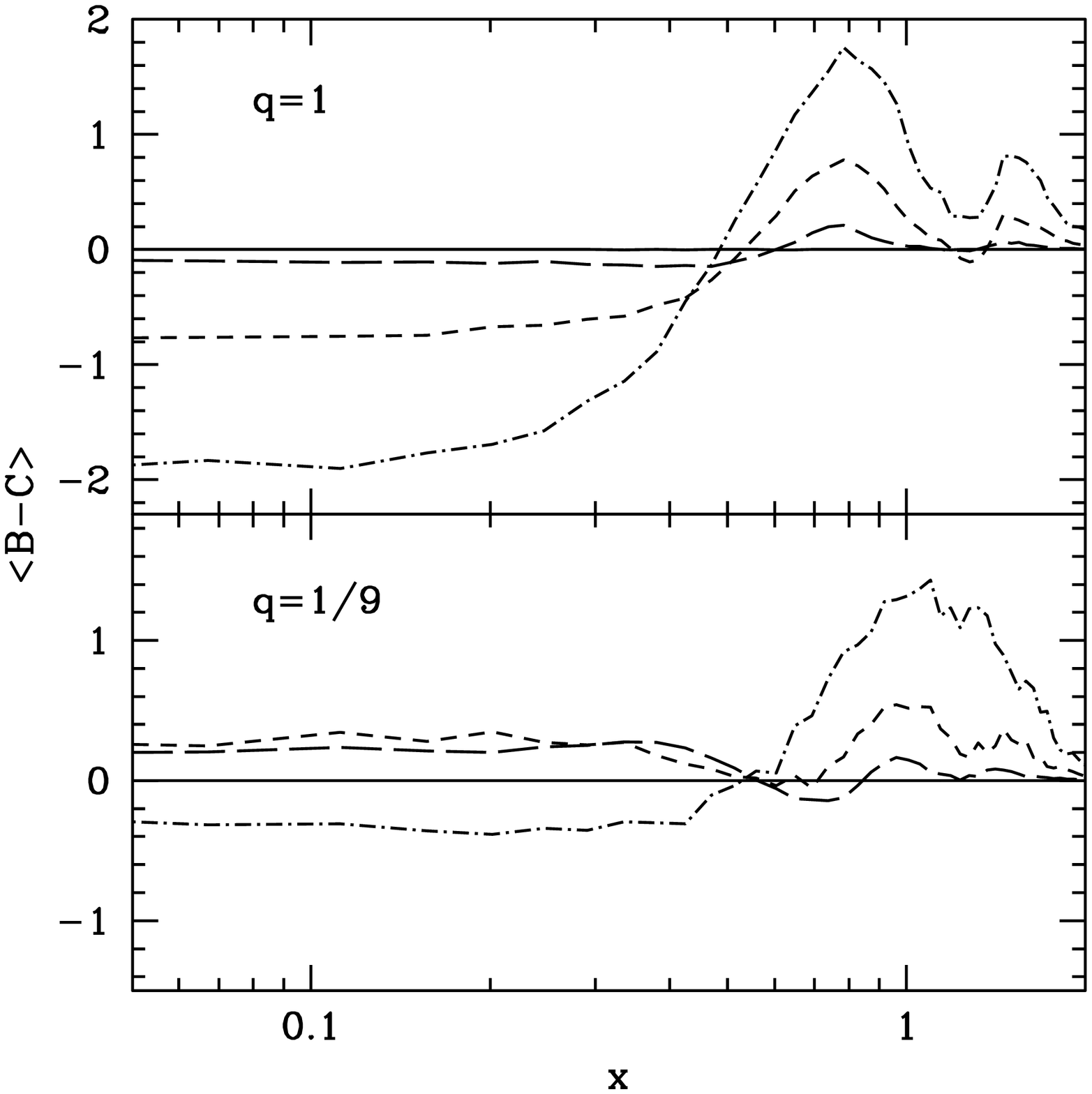,width=2.9in}}
\vspace{-0.0cm}
\caption{\footnotesize Mean angular momentum minus energy exchange 
$\langle B-C \rangle$
parameter as a function of dimensionless impact parameter $x$ for a binary
with eccentricity $e=0$ ({\it solid line}), 0.3 ({\it long-dashed line}), 0.6 
({\it short-dashed line}), and 0.9 ({\it dot-dashed line}). {\it Top panel}: 
equal-mass binary ($q=1$). {\it Bottom panel}: $q=1/9$.}
\label{fig:B-C}
\vspace{+0.5cm}
\end{figurehere}

Figure \ref{fig:C} shows examples of the energy exchange parameter 
$\langle C \rangle$ versus the star impact parameter $b$ 
in the limit $v\ll V_c$ i.e. for a ``hard'' binary. We 
have defined a rescaled dimensionless impact parameter $x$ as in Q96:
\begin{equation}\label{xpar}
x\equiv b/b_0,\,\,\,\,\, b_0^2=2GMa/v^2,
\end{equation}
where $b_0$ is the impact parameter {\it at infinity} leading to a star 
pericenter $r_p=a$ including gravitational focusing. For a circular, 
equal-mass binary, $\langle C \rangle$ has a maximum for impact parameter 
equal to the distance of each binary member from the center of mass
($x\simeq 0.5$), and then decreases reaching a constant value $\langle 
C\rangle \simeq 1$ for small impact parameters. In the case of highly 
eccentric binaries ($e=0.9$), the energy exchange is almost monotonic: 
a star with small impact parameter interacts with the pair near pericenter 
when it has maximum speed, and this results in a large energy exchange. 
For circular binaries and $x\lesssim 1$, one finds $1.5 \lesssim \langle 
C\rangle \lesssim 2$ nearly independently of the mass ratio. The peak at 
$x=1$ for $q\ll 1$ corresponds to the location of $M_2$, as $M_1$ is now 
at rest and $M_2$ orbits around $M_1$ at distance $a$. From equation 
(\ref{c}) it follows that stars typically experience an energy gain 
proportional to the reduced mass of the binary, and inversely proportional 
to the binary separation. The combination $\langle B(x)-C(x)\rangle$, which 
governs the eccentricity evolution (see eq. \ref{K1}), is shown in 
Figure \ref{fig:B-C}. For a circular binary $\langle B-C\rangle$ is zero 
independently of impact parameter, i.e. circular binaries remain circular. If 
the binary is eccentric, $\langle B-C \rangle$ shows a positive peak for 
impact parameters corresponding to close encounters with $M_2$ 
or with one member of an equal-mass pair. Interactions 
with $0.5\lesssim x\lesssim 2$ stars tend to increase the eccentricity of 
the binary, whereas stars with low impact parameter have either a small 
effect ($q=1/9$) or can significantly decrease the eccentricity ($q=1$).
Typically, $B\propto\Delta L_{z*}$ assumes positive values,
i.e. scattered stars gain angular momentum along the $z$-axis
and will tend to corotate with the binary. 

\section{Hardening in a fixed background}

As described in Q96, the binary evolution in an isotropic fixed background of stars 
of density $\rho$ and one-dimensional velocity dispersion $\sigma$ at infinity 
is determined by three dimensionless quantities: the hardening rate
\begin{equation}\label{dadtmaxw}
H={\sigma\over G\rho}\frac{d}{dt}\left(\frac{1}{a}\right),
\end{equation}
the mass ejection rate ($M_{\rm ej}$ is the stellar mass ejected by the
binary)
\begin{equation}\label{jsigma}
J=\frac{1}{M} \frac{dM_{\rm ej}} {d\ln(1/a)},
\end{equation}
and the eccentricity growth rate
\begin{equation}\label{de}
K=\frac{de}{d\ln(1/a)}.
\end{equation}
The average hardening rate for a Maxwellian stellar velocity distribution 
$f(v,\sigma)=(2\pi\sigma^2)^{-3/2}$ $\exp(-v^2/2\sigma^2)$ is 
\begin{equation}\label{hsigma}
H(\sigma) \equiv \int_0^{\infty} f(v,\sigma)\frac{\sigma}{v}H_1(v)\,4\pi v^2\,dv,
\end{equation}
where 
\begin{equation}\label{H1}
H_1(v) \equiv 8\pi\int_0^{\infty} \langle C\rangle x\,dx
\end{equation}
is the dimensionless hardening rate if all stars have the same velocity $v$. An 
expression analogous to equation (\ref{hsigma}) relates the thermally-averaged 
eccentricity growth rate $K(\sigma)$ to $K_1(v)$, where
\begin{equation}\label{K1}
K_1(v) \equiv \frac{(1-e^2)}{2e}\frac{\int_0^{\infty}\langle B-C\rangle x\,dx}
{\int_0^{\infty}\langle C\rangle x\,dx}.
\end{equation}
Both $H_1$ and $K_1$ are independent of $M$ and $m_*$. Figure~\ref{fig:Hsigma} shows 
the hardening rate $H$ versus binary separation $a$ as derived from our 
scattering experiments. As found by Q96, $H$ is approximatively constant for 
\begin{equation}\label{hardrad}
a<a_h=\frac{GM_2}{4\sigma^2}.
\end{equation}
Defining as ``hard'' a binary whose orbital separation is smaller than $a_h$, 
it follows then that a ``hard binary hardens at a constant rate''. Note that 
there is no explicit dependence on $\sigma$ once the rate is expressed as a function 
of $a/a_h$. 
$H$ is found to be a decreasing function of $q$ but increases with 
increasing eccentricity. The latter trend can be understood from Figure~\ref{fig:C}, 
which shows a significative enhancement in the energy change 
$\langle C \rangle$ at small impact parameters at increasing eccentricity. 
Note that $H$ drops to zero for $a\gtrsim 10\,a_h$.
We choose the following ejection criterion to measure the rate at which the binary ejects stars. 
The binary is assumed to be embedded in
a bulge of mass $M_B$ and stellar density profile
approximated by a singular isothermal sphere (SIS).
Stars are counted as ``ejected'' from the bulge if, after three-body
scattering, their velocity $V$ far away from the binary is greater
than the escape velocity from the radius of influence of the binary,
$r_{\rm inf}\equiv GM/(2\sigma^2)$. The
SIS potential is $\phi(r)=-2\sigma^2\,[\ln (GM_B/2\sigma^2r)+1]$
(for $r<R_B=GM_B/2\sigma^2$), and the escape speed from $r_{\rm inf}$ is then
\begin{equation}
v_{\rm esc}\equiv \sqrt{-2\phi}=2\sigma \sqrt{[\ln(M_B/M)+1]}=5.5\sigma, 
\label{eq:vesc}
\end{equation}
where the second equality comes from the adopted bulge--black hole mass relation 
$M=0.0014\,M_B$ (Haring \& Rix 2004). Note that this yields a more conservative
ejection criterion than the conventional choice $V>v_{\rm esc}=2\sqrt{3}\sigma$ 
adopted by Q96.
\footnote{When the binary first becomes hard, only a few stars
acquire a kick velocity large enough to escape the host bulge. Many 
scattered stars will instead return to the central region on nearly unperturbed, 
small impact parameter orbits, and will undergo a second super-elastic scattering 
with the binary. We will quantify the role of these ``secondary slingshots'' in 
determining the hardening of the pair in a subsequent work.}\
Denoting with $F_{\rm ej}(x,v,\sigma)$ the fraction of incident stars with impact 
parameter $x$ and initial velocity $v$ that satisfy equation (\ref{eq:vesc})  
after a three-body interaction, the thermally averaged ejection rate can 
then be written as  
\begin{equation}\label{Jsigma}
J(\sigma) \equiv {1\over H} \int_0^{\infty} f(v,\sigma)\,{\sigma\over v}\,4\pi v^2\,dv
\,4\pi\, \int_0^\infty F_{\rm ej}\,x\, dx.
\end{equation}
We have found that the (Maxwellian averaged) rates $H$, $J$, and $K$ derived 
from scattering experiments can be fitted to within few percent by the following 
functions:
\begin{equation}\label{fit}
H=A (1+a/a_0)^{\gamma},
\end{equation}
\begin{equation}\label{fit2}
J=A (a/a_0)^{\alpha}[1+(a/a_0)^{\beta}]^{\gamma},
\end{equation}
and
\begin{equation}\label{fitK}
K=A(1+a/a_0)^{\gamma}+B.
\end{equation}
The parameters of the fits to $H$ and $J$ are listed in Tables \ref{Tab1} and \ref{Tab2} 
for different binary mass ratios and circular orbits, while fit parameters to $K$ are 
listed in Table \ref{Tab3} for different values of $q$ and $e$. 
The binary eccentricity growth and mass ejection rates $K$ and $J$ (averaged over a 
Maxwellian distribution) are plotted in Figures \ref{fig:Ksigma} and \ref{fig:Jsigma} 
as a function of $a/a_h$. The parameter $K$ is close to zero for $a\sim a_h$, and grows 
monotonically in the case of eccentric orbits as the binary shrinks. Smaller values 
of $q$ lead to larger growth rates. In the circular case $K$ is negligible at 
every separation, i.e. circular binaries remain circular. The mass ejection rate 
has only a weak dependence on eccentricity and mass ratio.

\begin{tablehere}
\begin{center}
\begin{tabular}{cccc}
\hline
$q$  & $A$ & $a_0$ & $\beta$\\
\hline
   1&       14.55&     3.48&     0.95\\
   1/3&      15.82&     4.18&     0.90\\
   1/9&      17.17&     3.59&     0.79\\
   1/27&     18.15&     3.32&     0.77\\
   1/81&     18.81&     3.87&     0.82\\
   1/243&    19.16&     4.16&     0.86\\
\hline
\end{tabular}
\end{center}
\caption{\footnotesize Best fit parameters describing (see eq.~\ref{fit}) 
the hardening rate $H$ for 
a circular binary with varying mass ratio $q$. The parameter $a_0$ is 
given in units of $a_h$.}
\label{Tab1}
\end{tablehere}
\begin{tablehere}
\begin{center}
\begin{tabular}{cccccc}
\hline
$q$  & $A$ & $a_0$ & $\gamma$ & $\alpha$ & $\beta$\\
\hline
   1&       0.224&     1.741&   -10.986&  -0.165& 1.095\\
   1/3&     0.201&     1.784&    -7.360&  -0.185& 1.176\\
   1/9&     0.214&     0.803&    -2.738&  -0.200& 1.291\\
   1/27&    0.215&     0.565&    -1.853&  -0.207& 1.374\\
   1/81&    0.211&     0.424&    -1.319&  -0.220& 1.564\\
   1/243&   0.210&     0.389&    -1.210&  -0.222& 1.638\\
\hline
\end{tabular}
\end{center}
\caption{\footnotesize Best fit parameters describing (see eq.~\ref{fit2}) the mass 
ejection rate $J$ for a circular binary with varying mass ratio. The parameter 
$a_0$ is given in units of $a_h$.}
\label{Tab2}
\end{tablehere}
\begin{tablehere}
\begin{center}
\begin{tabular}{cccccc}
\hline
$q$  & $e$ & $A$ & $a_0$ & $\gamma$ & $B$\\
\hline
   &   0.15&    0.037&     0.339&   -3.335&   -0.012\\
   &   0.3&     0.075&     0.151&   -1.548&   -0.008\\
   1&   0.45&   0.105&     0.088&   -0.893&   -0.005\\
   &   0.6&     0.121&     0.090&   -0.895&   -0.008\\
   &   0.75&    0.134&     0.064&   -0.544&   -0.006\\
   &   0.9&     0.082&     0.085&   -0.663&   -0.004\\
\hline
   &      0.15&    0.082&     0.042&   -0.168&   -0.048\\
   &       0.3&    0.095&     0.213&   -1.152&   -0.012\\
   1/3&   0.45&    0.129&     0.137&   -0.655&   -0.006\\
   &       0.6&    0.166&     0.081&   -0.546&   -0.006\\
   &      0.75&    0.159&     0.079&   -0.497&   -0.010\\
   &       0.9&    0.095&     0.122&   -0.716&   -0.008\\
\hline
   &      0.15&    0.051&     0.385&   -0.891&   -0.011\\
   &       0.3&    0.111&     0.307&   -1.107&   -0.007\\
   1/9&   0.45&    0.172&     0.526&   -1.174&   -0.016\\
   &       0.6&    0.181&     0.251&   -1.169&   -0.007\\
   &      0.75&    0.179&     0.195&   -0.846&   -0.004\\
   &       0.9&    0.117&     0.400&   -1.170&   -0.001\\
\hline
   &      0.15&    0.064&     0.284&   -1.206&   0.021\\
   &       0.3&    0.143&     1.033&   -1.537&   -0.021\\
   1/27&  0.45&    0.212&     0.722&   -1.257&   -0.022\\
   &       0.6&    0.216&     0.430&   -1.163&   -0.014\\
   &      0.75&    0.173&     0.771&   -1.934&   -0.014\\
   &       0.9&    0.129&     0.329&   -1.125&   -0.020\\
\hline
\end{tabular}
\end{center}
\caption{\footnotesize Best fit parameters describing (see eq.~\ref{fitK}) the 
eccentricity growth rate $K$ for varying initial eccentricity $e$ and mass 
ratio $q$. The parameter $a_0$ is given in units of $a_h$.}
\label{Tab3}
\end{tablehere}
It is interesting to compare our fits to those provided by Q96. While our results for 
the dimensionless rates $H_1$ and $K_1$ are consistent, within the 
statistical errors, to those of Q96, we have fit directly the Maxwellian-averaged 
rates for ease of use. Our choice of a different velocity threshold for ejection 
makes our mass ejection rate $J(\sigma)$ generally lower than that of Q96. 
Regarding $H(\sigma)$, limiting values for small separations are virtually identical
to those of Q96, while at $a\sim a_h$ the fit in Q96 is $20$\% above our 
formula. As for the eccentricity growth rate, we have covered a much wider region of the 
$(q,e)$ parameter space, compared to the limited sampling provided by Q96.
\begin{figurehere}
\vspace{0.5cm} 
\centerline{\psfig{figure=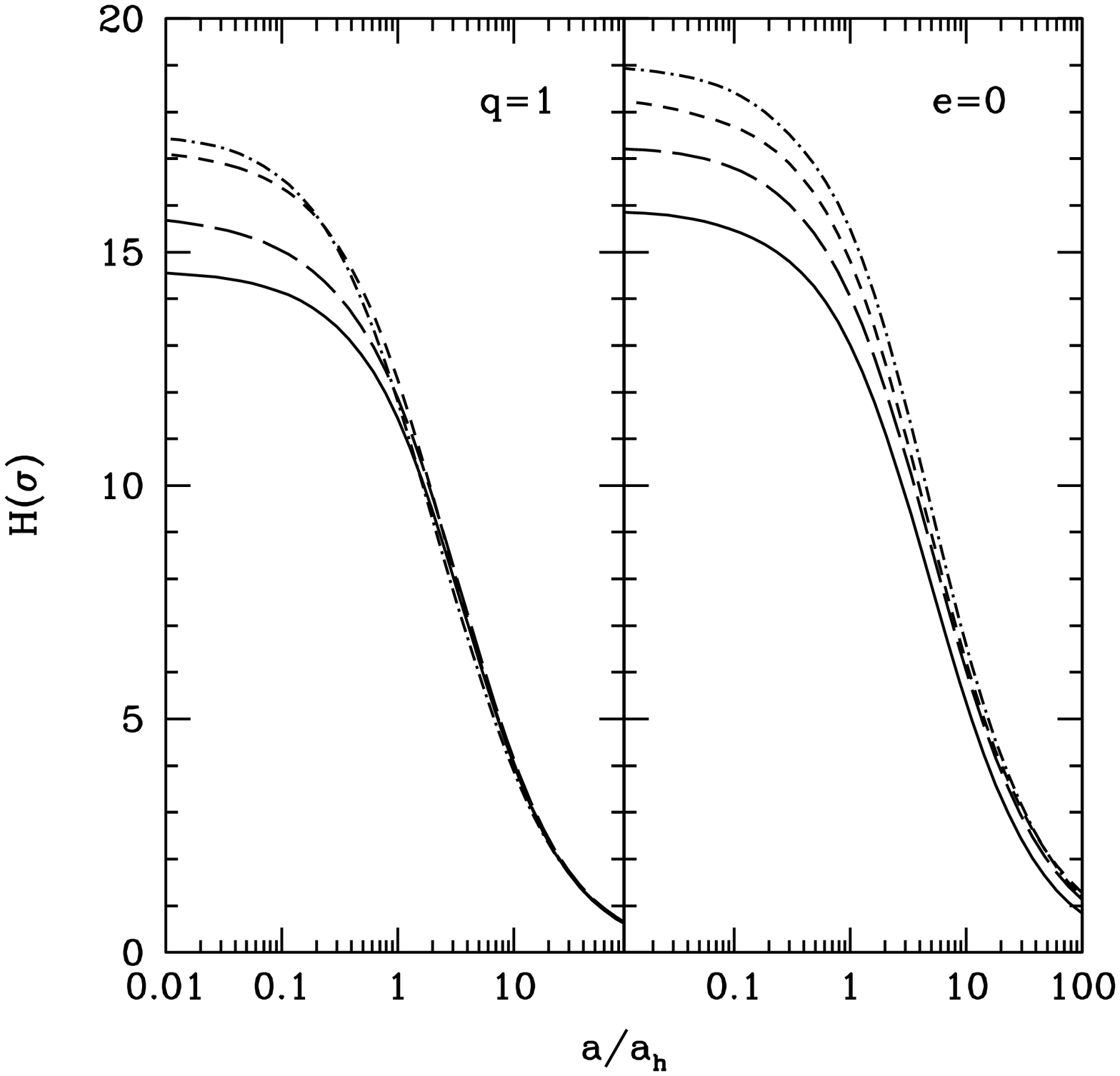,width=2.9in}}
\vspace{-0.0cm}
\caption{\footnotesize Binary hardening rate $H$ (averaged over a Maxwellian velocity distribution) 
versus $a/a_h$. {\it Left panel:} $e=$0, 0.3, 0.6, 0.9 for $q=1$.  
{\it Right panel:} $q=$1/3, 1/9, 1/27, 1/81 for $e=0$. 
Line style as in Fig.~\ref{fig:C}.}
\label{fig:Hsigma}
\vspace{+0.5cm}
\end{figurehere}
\begin{figurehere}
\vspace{0.5cm} 
\centerline{\psfig{figure=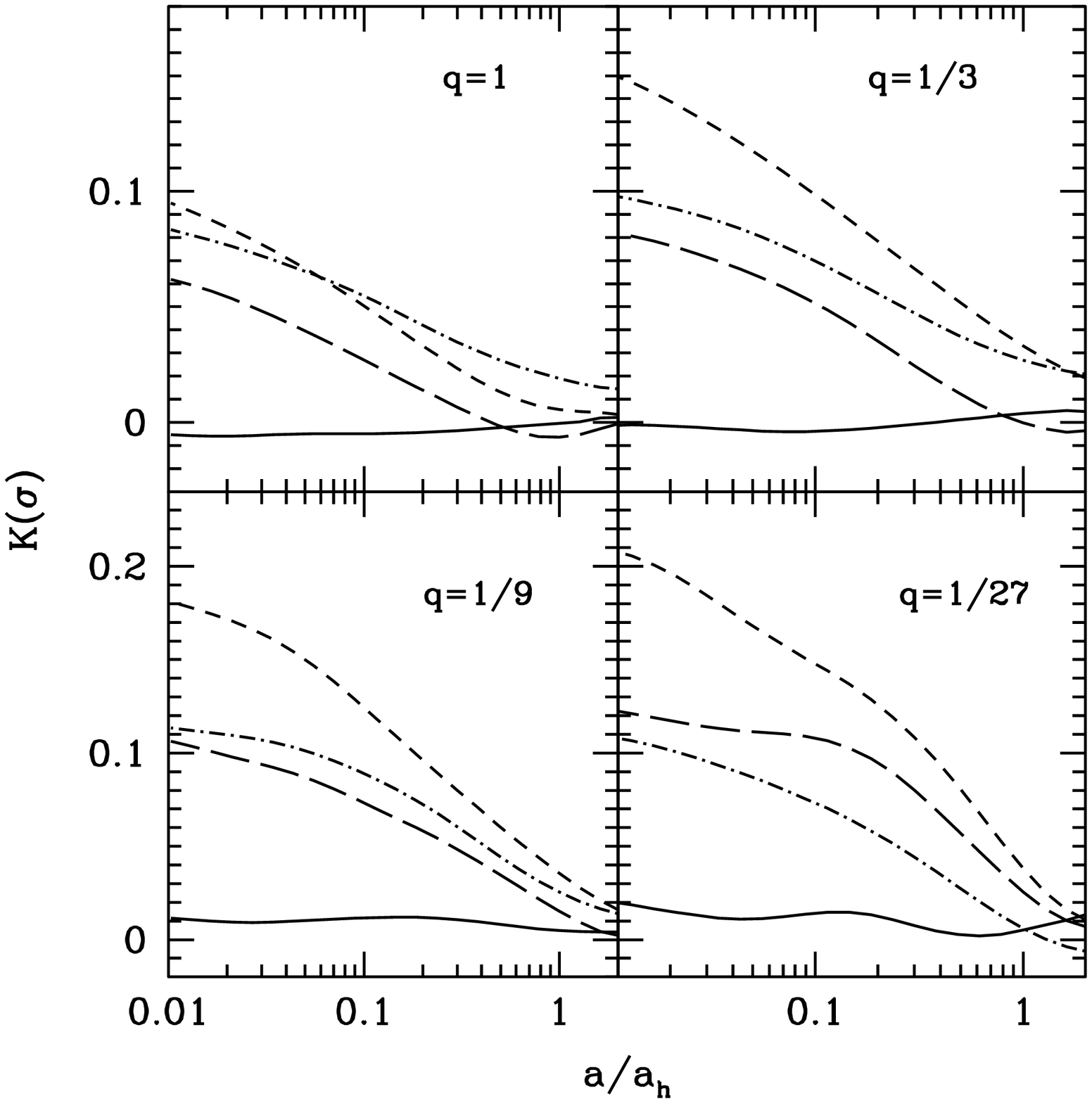,width=2.9in}}
\vspace{-0.0cm}
\caption{\footnotesize Binary eccentricity growth rates as a function of binary 
separation for different mass ratios. Different line styles are for $e=$0, 0.3, 
0.6, 0.9, as in the top panel of Fig.~\ref{fig:C}.}
\label{fig:Ksigma}
\vspace{+0.5cm}
\end{figurehere}

\section{Kinematics of hypervelocity stars}

When the MBHB separation is $a\lesssim a_h$, 
only a small fraction ($\lesssim 1$\%) of bulge stars have low-angular momentum 
trajectories with pericenters lying within $a$ (the binary's geometrical 
``loss cone''). If the loss-cone is constantly refilled as
the pair shrinks, then a substantial subpopulation of suprathermal HVSs will be produced
via the gravitational slingshot. In this section we use our scattering experiments 
to study the kinematic of HVSs as a function of binary mass ratio, eccentricity, 
and orbital separation.   
\begin{figurehere}
\vspace{0.5cm} 
\centerline{\psfig{figure=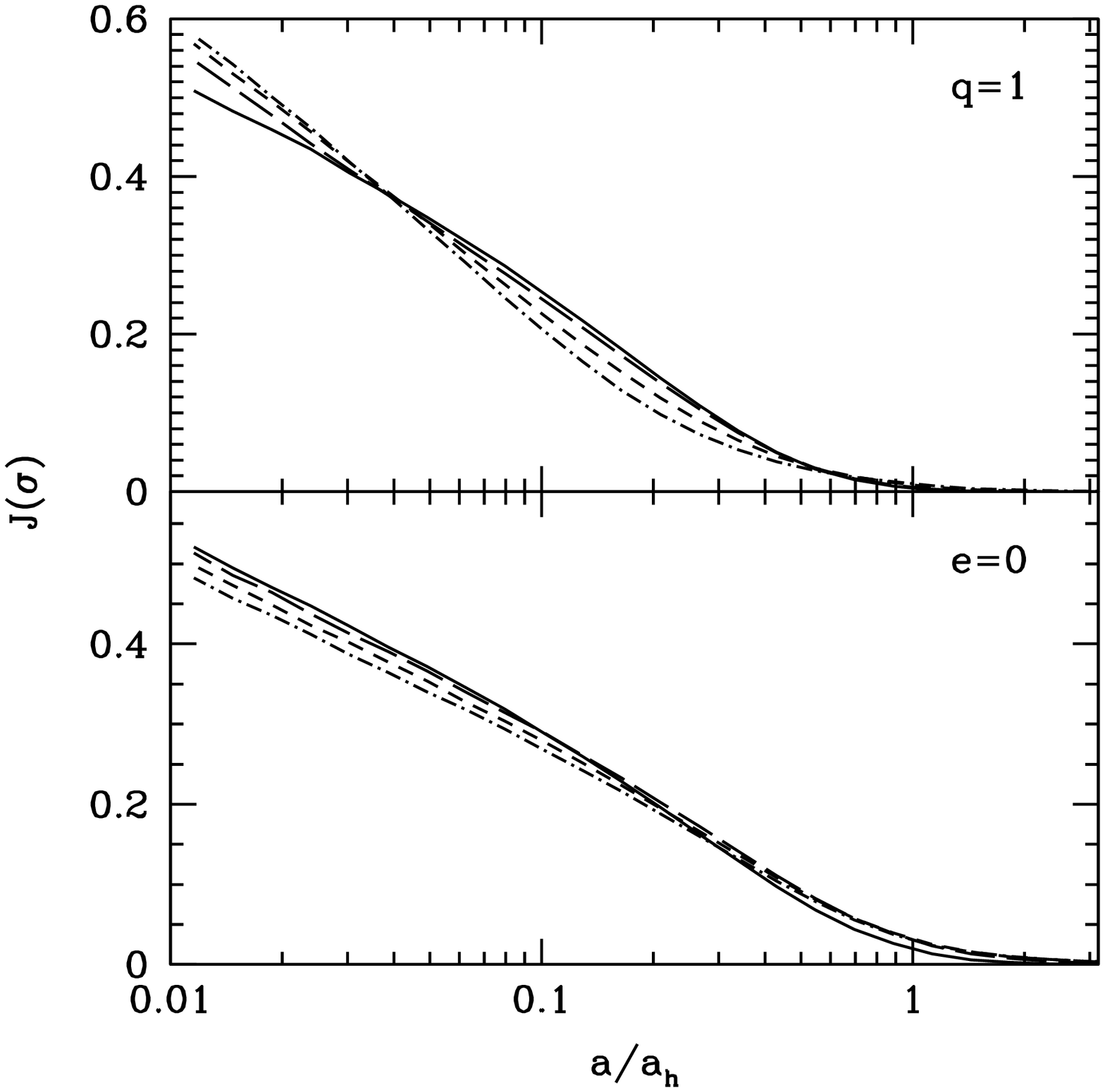,width=2.9in}}
\vspace{-0.0cm}
\caption{\footnotesize Mass ejection rate $J$ versus
$a/a_h$. {\it Top panel:} $e=$0, 0.3, 0.6, 0.9 for $q=1$.  
{\it Bottom panel:} $q=$1/3, 1/9, 1/27, 1/81 for $e=0$. 
Line style as in Fig.~\ref{fig:C}.}
\label{fig:Jsigma}
\vspace{+0.5cm}
\end{figurehere}

\subsection{Velocity distribution}

In a scattering event, a star that starts with a low initial velocity $v$ 
passes the two MBHs at a distance $\sim a$ and leaves with a gain to its kinetic 
energy, $V^2\simeq 2C(\mu/M)V_c^2=[8C\sigma^2/(1+q)](a_h/a)$ (eq.~\ref{c}), 
where $C$ depends on the impact parameter (see Fig.~\ref{fig:C}). 
Figure \ref{fig:vmean} shows the average final velocity of scattered stars $\langle V
\rangle$ as a function of inverse binary separation, for different mass ratios
and eccentricities. The population includes all stars with maximum impact parameter 
at infinity  $x\leq 2$, corresponding to a pericenter $r_p\leq 4a$. 
The curves clearly follow the expected $V\propto \sqrt{a_h/a}$ scaling. For
$a_h/a \gtrsim 10$, most scattering events produce HVSs that escape from the 
bulge. Neither the binary mass ratio nor its eccentricity have a large effect 
on the average final velocity. Larger eccentricities produce a more prominent 
high-velocity tail of HVSs as in this case the orbital velocity of $M_2$ 
close to pericenter is larger than $V_c$, allowing for more energetic 
slingshots. Our numerical experiments show that both a small mass ratio and a 
large eccentricity tend to increase the fraction of extremely energetic 
scattering events.

The properties of scattered stars are best described noticing that, when stellar 
ejection velocities are measured in units of binary orbital velocity $V_c$, 
the high-velocity tail of the resulting distribution function is 
actually scale-invariant, i.e. independent of binary separation for fixed 
$q$ and $e$. This is clearly shown in Figure~\ref{fig:vesc1}. Scale 
invariance is broken by the choice of an absolute velocity threshold 
(e.g. $V>v_{\rm esc}$) that, in units of $V_c$, is a decreasing function 
of $a_h/a$. The high-velocity tail of the distribution function of scattered 
stars can be described as a broken power-law in the range $4\sigma \lesssim 
V\lesssim 3 V_{\rm max}$, where $V_{\rm max}=V_c\,\sqrt{(1+e)/
(1-e)}/(1+q)$ is the velocity of the lighter black hole at the pericenter:
\begin{equation}\label{eq:fvel}
f(w)=\frac{A}{h}(w/h)^{\alpha}\left[1+(w/h)^{\beta}\right]^{\gamma},
\end{equation}
where $w\equiv V/V_c$, and $h\equiv \sqrt{2q}/(1+q)$. Best fitting parameters 
are given in in Table~\ref{Tab4} for different values of the eccentricity, 
while the fitting formula and the scattering experiment data are 
compared in Figure~\ref{fig:vesc2}. The velocity distribution of 
HVSs ($V>v_{\rm esc}$) at a given binary separation is related to 
the mass ejection rate by 
\begin{equation}\label{eq:Jandfvel}
J(a)=-\frac{d(M_{\rm ej}/M)}{da/a}=\int_{v_{\rm esc}/V_c}^{\infty}\,f(w)dw.
\end{equation}
Equation (\ref{eq:Jandfvel}) sets the normalization constant $A$ in equation 
(\ref{eq:fvel}). 

\begin{tablehere}
\begin{center}
\begin{tabular}{ccccc}
\hline
$e$  & $A$ & $\alpha$ & $\beta$ & $\gamma$\\
\hline
   0&         0.236&     -0.917&     16.365& -0.165\\
   0.3&       0.242&     -1.067&     11.722& -0.235\\
   0.6&       0.385&     -0.765&     4.627&  -0.726\\
   0.9&       0.556&     -0.599&     2.375&  -1.420\\
\hline
\end{tabular}
\end{center}
\caption{\footnotesize Best fit parameters describing (see eq.~\ref{eq:fvel}) the 
velocity distribution of scattered stars with $4 \sigma\lesssim V\lesssim 3V_{\rm max}$.}
\label{Tab4}
\end{tablehere}
\begin{figurehere}
\vspace{0.5cm} 
\centerline{\psfig{figure=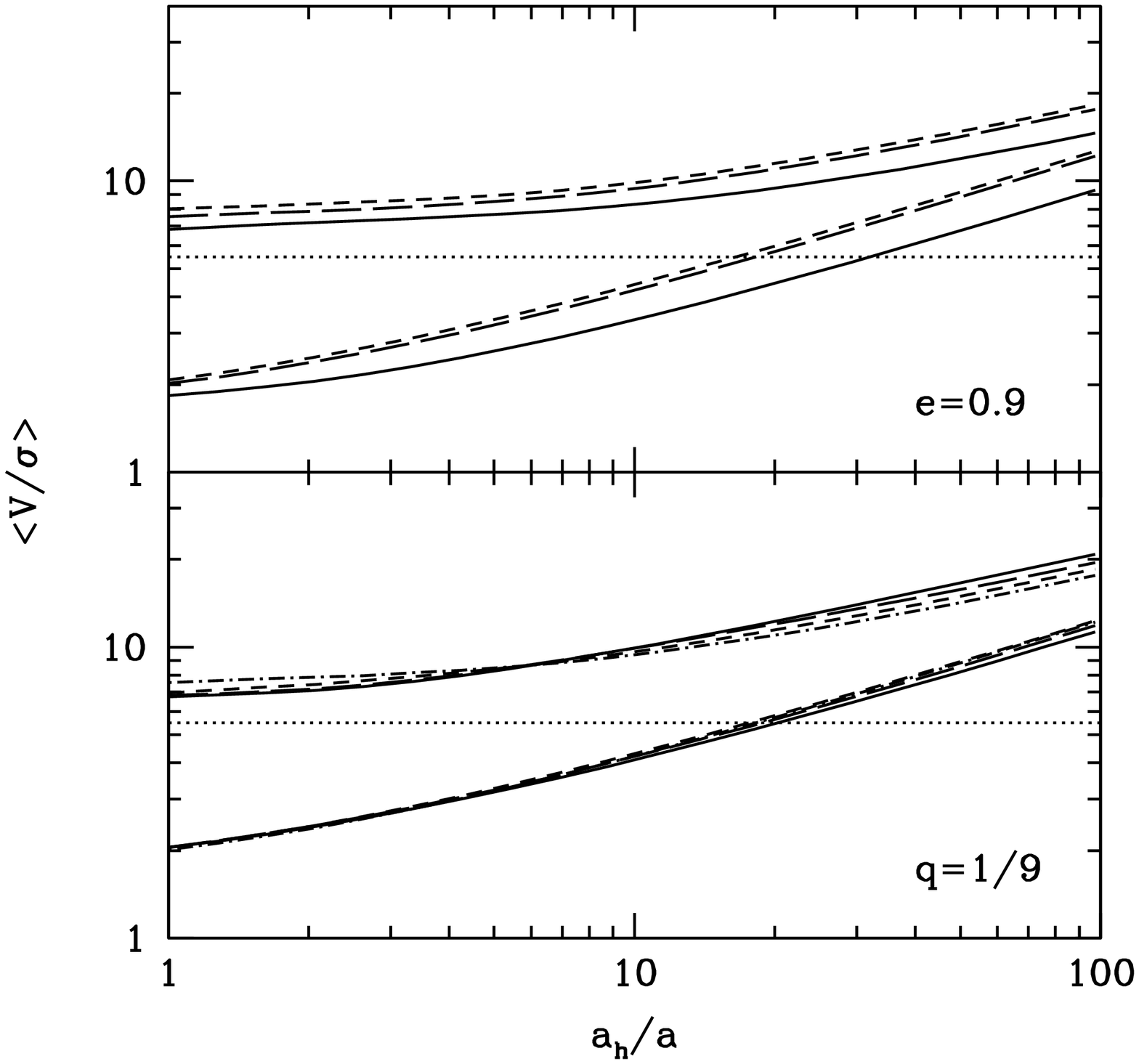,width=2.9in}}
\vspace{-0.0cm}
\caption{\footnotesize Average final velocity $\langle V\rangle$ of all scattered stars versus 
inverse binary separation $a_h/a$. {\it Top panel:} eccentric $e=0.9$ binary 
with $q=1$ ({\it solid lines}), $q=1/9$ ({\it long-dashed lines}), and $q=1/81$ ({\it 
short-dashed lines}). The lower set of curves shows results for the entire population of 
scattered stars (defined as those with impact parameter at infinity $x\leq 2$), 
while the upper 
set of curves shows only HVSs with $V>v_{\rm esc}=5.5\sigma$. The dotted line 
in both panels marks the value $V=v_{\rm esc}$. {\it Bottom panel:} same but for $q=1/9$ 
binary with $e=0$ ({\it solid lines}), $e=0.3$ ({\it long-dashed lines}), $e=0.6$ 
({\it short-dashed lines}), and $e=0.9$ ({\it dot-dashed lines}).}
\label{fig:vmean}
\vspace{+0.5cm}
\end{figurehere}
\begin{figurehere}
\vspace{0.5cm} 
\centerline{\psfig{figure=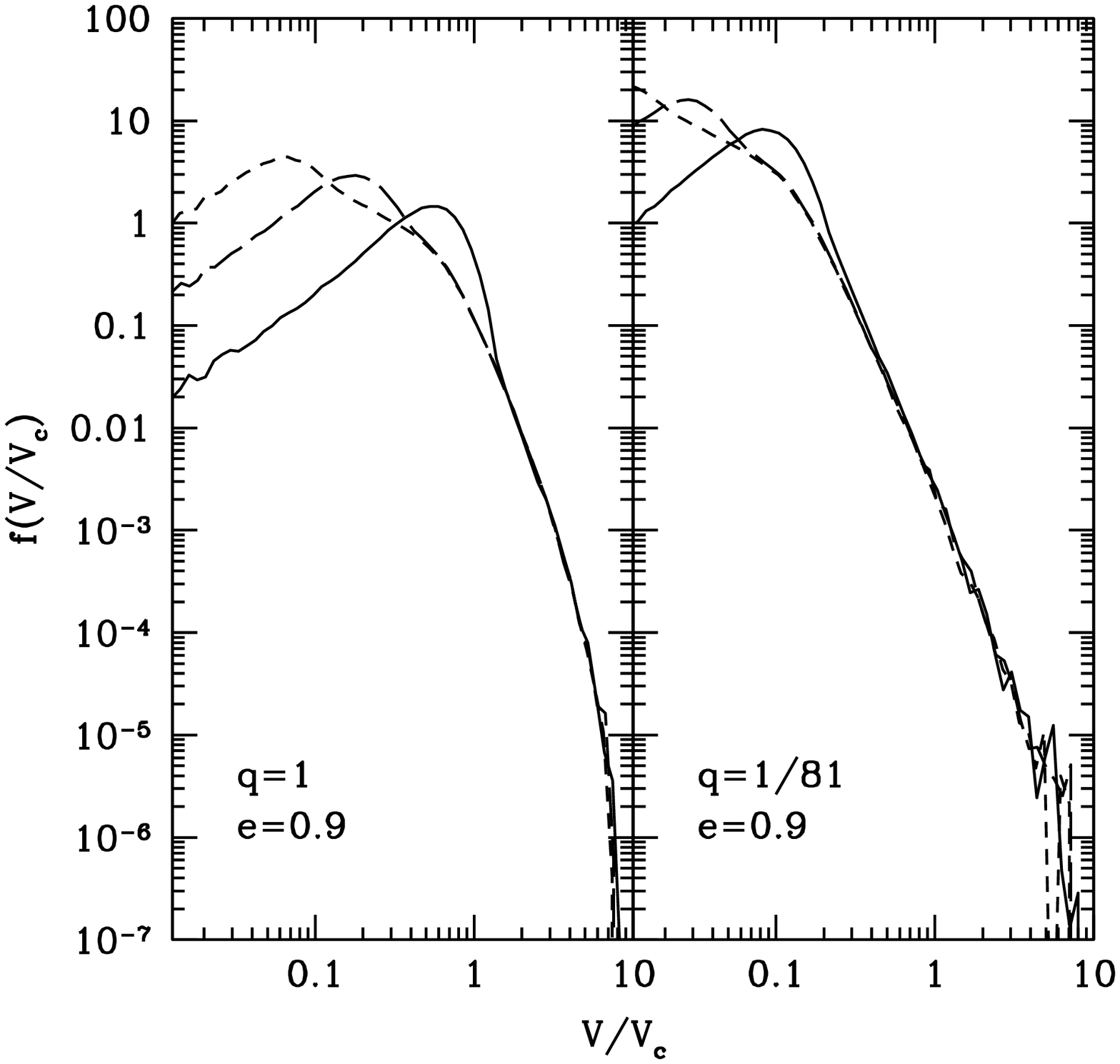,width=2.9in}}
\vspace{-0.0cm}
\caption{\footnotesize Differential distribution of $V/V_c$ for all stars scattered at binary 
separation $a_h/a=1$ ({\it solid line}), 10 ({\it long-dashed line}), and 100 ({\it 
short-dashed line}). As the binary shrinks the peak shifts to increasingly small 
values of $V/V_c$ as injected stars are drawn from a thermal distribution of 
fixed velocity dispersion $\sigma$. {\it Left panel:} binary with $q=1$ and $e=0.9$. 
{\it Right panel:} binary with $q=1/81$ and $e=0.9$.}
\label{fig:vesc1}
\vspace{+0.5cm}
\end{figurehere}
\begin{figurehere}
\vspace{0.5cm} 
\centerline{\psfig{figure=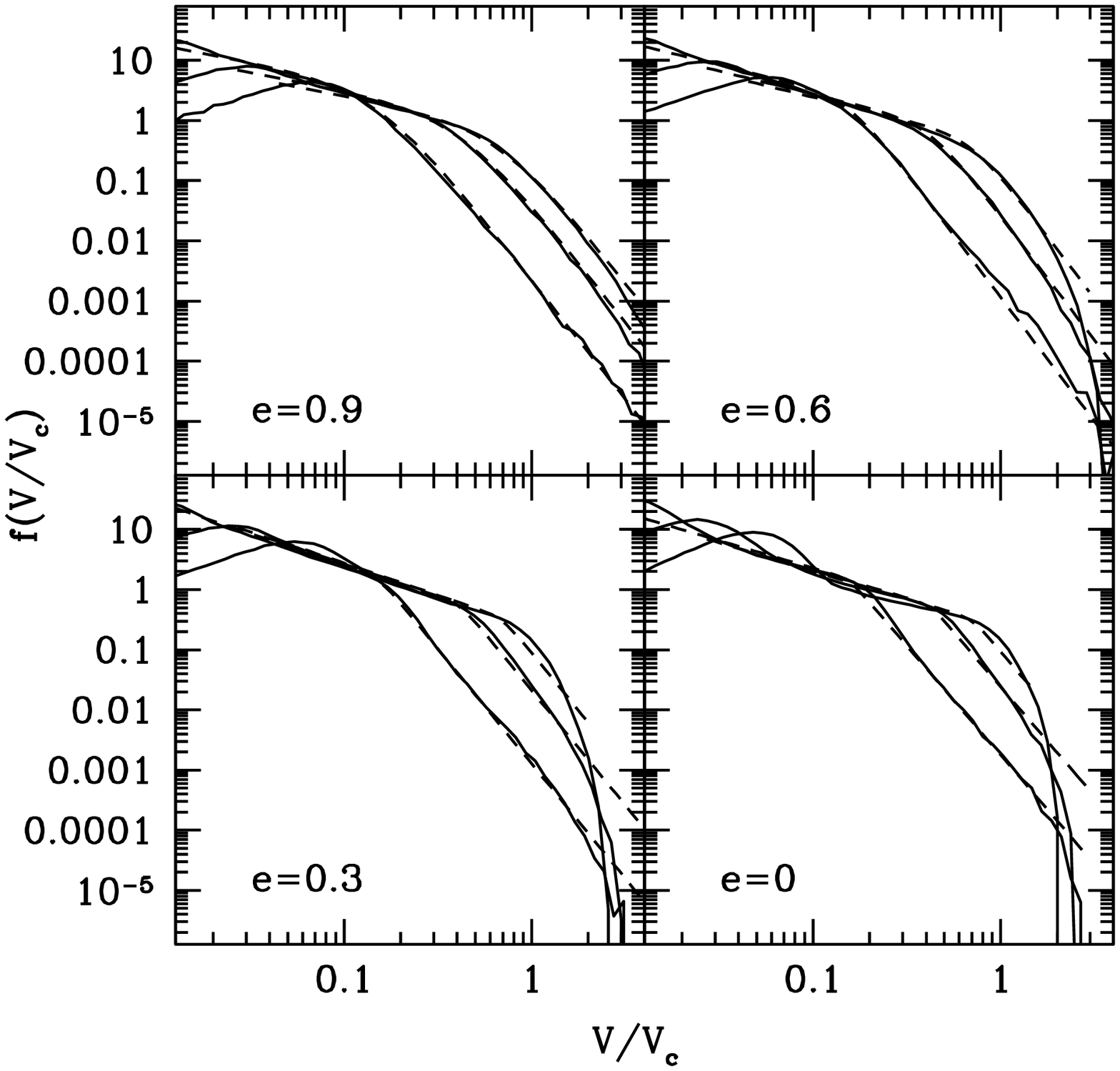,width=2.9in}}
\vspace{-0.0cm}
\caption{\footnotesize Differential distribution of $V/V_c$ for all stars scattered at binary
separation $a_h/a=100$. In each panel values are shown, right to left, for binary mass ratios
$q=1, 1/9$ and 1/81. Scattering experiment data are shown as solid lines, while 
fitting formula results, extended up to $V=3V_{\rm max}$ (eq.~\ref{eq:fvel}, and Table~\ref{Tab4}), as dashed lines.   
The appropriate values of the eccentricity are labeled in each panel.}
\label{fig:vesc2}
\vspace{+0.5cm}
\end{figurehere}
\subsection{Angular distribution}

Analytic expressions for the time-dependent phase-space distribution of stars ejected
from the Galactic center as a result of inspiral of an intermediate-mass black hole
($q\ll 1$) have been recently derived by Levin (2005). HVSs are found to follow a 
characteristic angular pattern: i) they are ejected preferentially in the orbital 
plane of the MBHB, and ii) perpendicular to the semimajor axis of the pair in the case of eccentric binaries.
For a given mass ratio and eccentricity, the 
magnitude of these three effects is a function of orbital separation and stellar 
final velocity. Energetic three-body
interactions eject stars in the direction of maximum black hole orbital speed,
generating a larger anisotropy in higher velocity stars, and a positive $z-$component 
of the stellar angular momentum. For small binary separations the degree of anisotropy 
of the ejected stars is reduced as, in order to generate a kick above a given $V$, 
the three-body interaction needs not to be as strong as in the case of large separation.

\subsubsection{Latitude} 

Figure \ref{fig:theta2} shows $\langle \theta^2 \rangle$ as a function of binary 
separation, where $-\pi/2<\theta<\pi/2$ is the latitude of ejected stars, 
i.e. the angle between the 
star velocity vector at infinity and the binary orbital plane. As an isotropic distribution 
would yield $\langle \theta^2 \rangle=\pi^2/4-2=0.467$, values lower than this indicates that 
the ejected stars tend to be flattened towards the binary orbital plane. We plot results 
for the population of HVSs as a whole ($V>v_{\rm esc}$) and for a subset with 
$V>9 v_{\rm esc}$. Note that a value $\langle \theta^2 \rangle=0$ at large 
separations simply 
means that no stars are ejected with a velocity exceeding the given threshold, not that 
stars are actually scattered exactly in the binary orbital plane. Higher velocity stars 
are more flattened into the orbital plane, and smaller binary separations lead to 
a more isotropic angular distributions. At a fixed mass ratio and orbital separation,
more eccentric binaries eject stars in a more isotropic fashion, while at a fixed 
eccentricity binaries with smaller mass ratios produces a more isotropic distribution of 
HVSs. 
 
\begin{figurehere}
\vspace{0.5cm} 
\centerline{\psfig{figure=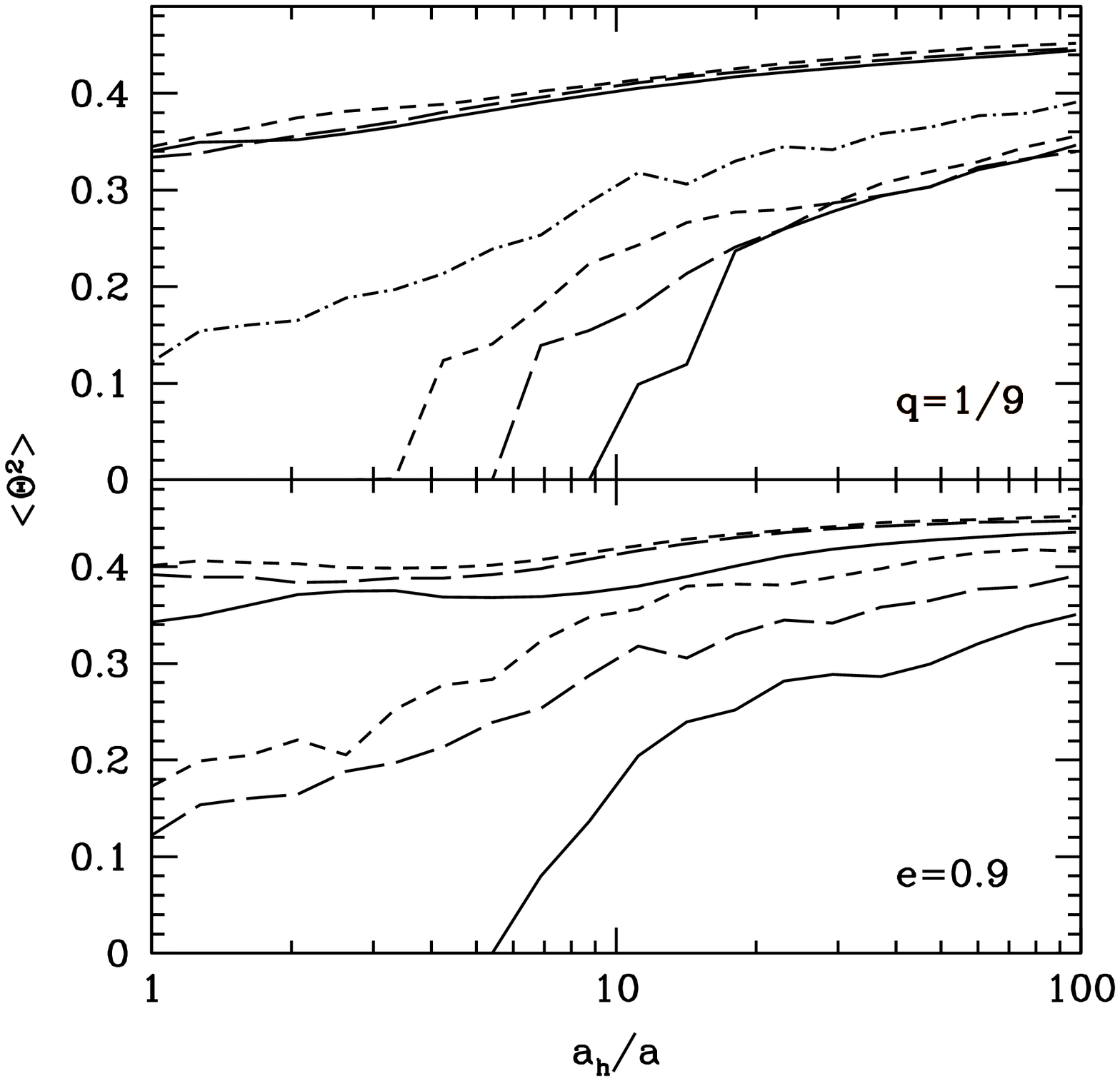,width=2.9in}}
\vspace{-0.0cm}
\caption{\footnotesize Mean value of $\theta^2$ versus $a_h/a$ for HVSs,
where $\theta$ is the latitude angle between the star velocity vector at infinity and 
the binary orbital plane. In each panel, the upper set of curves refers to all 
ejected stars ($V>v_{\rm esc}$), the lower set of curves to stars with 
$V>9v_{\rm esc}$. {\it Top panel}: binary with mass ratio $q=1/9$ and 
eccentricity $e=0$ ({\it solid lines}), 
$e=0.3$ ({\it long-dashed lines}), $e=0.6$ ({\it short-dashed lines}), and $e=0.9$ 
({\it dot-dashed lines}). {\it Bottom panel}: eccentric binary with $e=0.9$ mass 
ratio $q=1$ ({\it solid lines}), $q=1/9$, ({\it long-dashed lines}), and $q=1/81$ 
({\it short-dashed lines}).}
\label{fig:theta2}
\vspace{+0.5cm}
\end{figurehere}
\subsubsection{Longitude}

Scattered stars typically receive a kick along the direction of maximum velocity of the 
MBHs. This implies that, in the case of eccentric binaries, the spatial distribution of 
HVSs will form a broad jet perpendicular to the semimajor axis. For small binary mass 
ratios, the jet tends to be one-sided since $M_1$ is practically at rest and the 
interaction takes place close to the pericenter of $M_2$ (Levin 2005). For equal-mass 
binaries, we expect a two-sided symmetric jet instead. To quantify these effects we have
assumed that the semimajor axis coincides with the $x-$axis, and that the two holes
orbit counterclock-wise in the $xy$-plane. The $M_2$ ($M_1$) velocity vector at 
the pericenter then forms an angle $3\pi/2$ ($\pi/2$) relative to the $x-$axis. Let 
$\phi$ be the star longitude, i.e. the angle between the $x-$axis and the projection
onto the orbital plane of the star velocity vector at infinity. In the case of an
azimuthally symmetric distribution, $\langle \phi \rangle=\pi$, with relative 
dispersion $\sigma_{\phi}/\langle \phi \rangle=1/\sqrt{3}=0.58$. Figure 
\ref{fig:phie09} shows the mean longitude of HVSs and its dispersion for 
an eccentric binary with different mass ratios, as a function of $a_h/a$,
both for the whole population of HVSs stars ($V>v_{\rm esc}$) and for a a subset 
with $V>9\,v_{\rm esc}$. In the case $q\ll 1$ the jet is oriented along the velocity 
of $M_2$ at pericenter, while for $q=1$ the value $\langle \phi
\rangle\simeq \pi$ does not denote axisymmetry but rather a two-sided jet 
oriented perpendicular to the semimajor axis (the bottom panel shows 
that the angular dispersion is lower than its symmetric value). HVSs are ejected more 
symmetrically as the binary shrinks and, at any given separations, the degree of 
axisymmetry increases with decreasing kick velocity. The mean longitude angle 
as a function of eccentricity is shown in Figure \ref{fig:phiq9} for a MBHB binary 
with $q=1/9$. While the population of HVSs as a whole is nearly symmetric for 
all eccentricities regardless of binary separation, the high velocity subsample 
shows a clear azimuthal asymmetry for $a_h/a\lesssim 20$ that decreases
with decreasing separations. MBHBs already significantly beam HVSs for $e=0.3$: 
larger eccentricities produce similar values of $\langle \phi \rangle$ but at 
larger separations.

We note that, as well as ejection velocity, also the angular properties of 
scattered stars depend on the ratio $V/V_c$ but not on the binary
hardness $a_h/a$. This can be clearly seen in Figure \ref{fig:thetafivsVc},
where the mean latitude and longitude angles are plotted versus $V/V_c$ 
for a binary with $q=1/9$ and $e=0.6$, at separations $a/a_h=1, 0.1$ and 0.01.

\begin{figurehere}
\vspace{0.5cm} 
\centerline{\psfig{figure=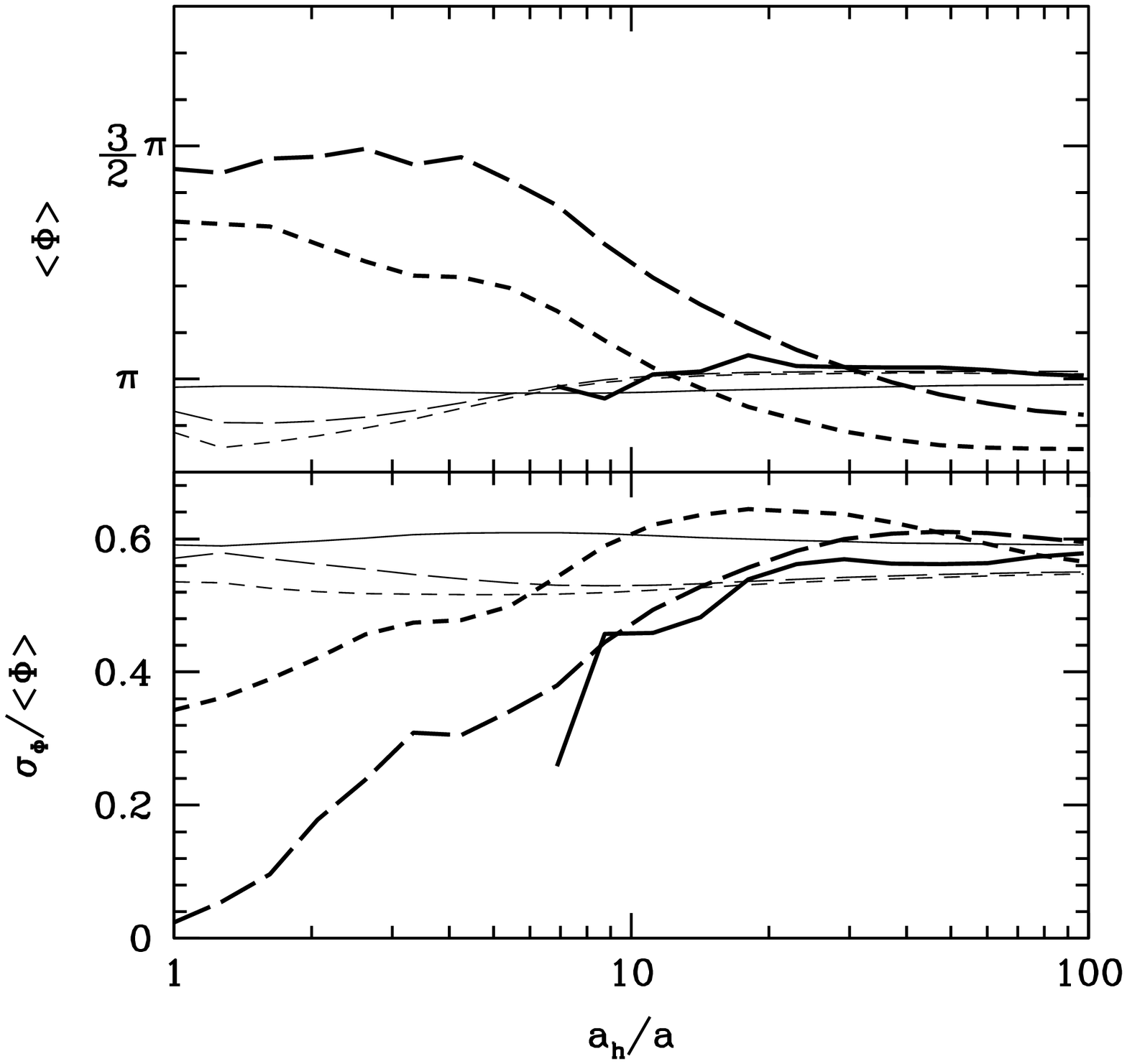,width=2.9in}}
\vspace{-0.0cm}
\caption{\footnotesize {\it Top panel:} mean longitude $\langle \phi \rangle$ of ejected
stars as a function of orbital separation. The binary semimajor axis lies along the 
$x-$vector, the MBHs orbit counterclock-wise in the $xy-$plane, and the binary 
eccentricity is set to 0.9. The solid, short-dashed, and long-dashed curves show
values for $q=1, 1/9$, and 1/81, respectively. {\it Thin lines:} entire population 
of ejected stars ($V>v_{\rm esc}$). {\it Thick lines:} subset of HVSs with 
$V>9\,v_{\rm esc}$. {\it Bottom panel:} relative azimuthal dispersion 
$\sigma_{\phi}/\langle \phi\rangle$. 
}
\label{fig:phie09}
\vspace{+0.5cm}
\end{figurehere}
\begin{figurehere}
\vspace{0.5cm} 
\centerline{\psfig{figure=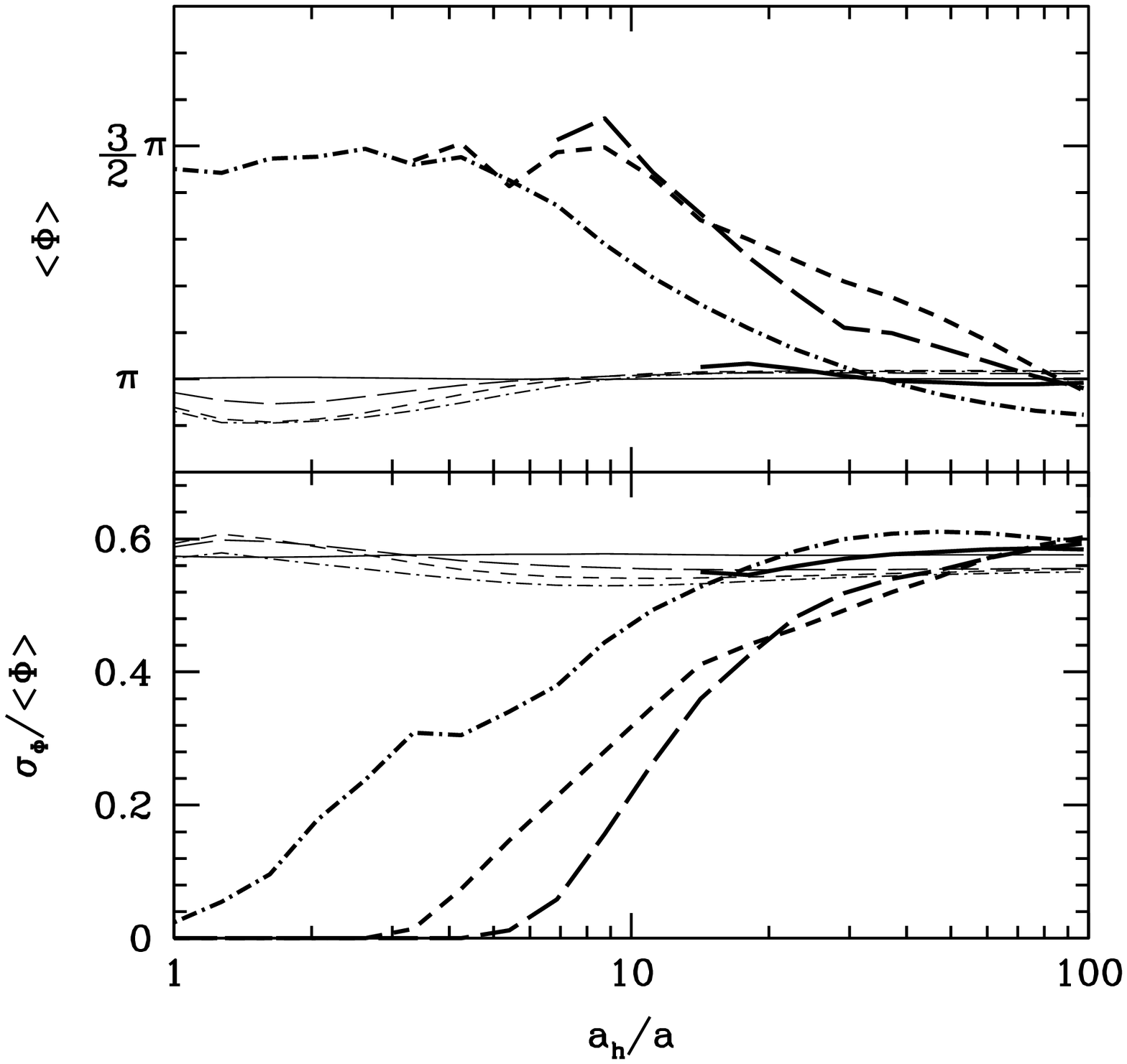,width=2.9in}}
\vspace{-0.0cm}
\caption{\footnotesize Same as Fig.~\ref{fig:phie09} but for a binary with $q=1/9$ and 
eccentricity $e=0$ ({\it solid lines}), $0.3$ ({\it long-dashed lines}), $0.6$ ({\it 
short-dashed lines}), and $0.9$ ({\it dot-dashed lines}). {\it Thin lines:} entire 
population of ejected stars ($V>v_{\rm esc}$). {\it Thick lines:} subset of HVSs 
with $V>9\,v_{\rm esc}$.}
\label{fig:phiq9}
\vspace{+0.5cm}
\end{figurehere}
\begin{figurehere}
\vspace{0.5cm} 
\centerline{\psfig{figure=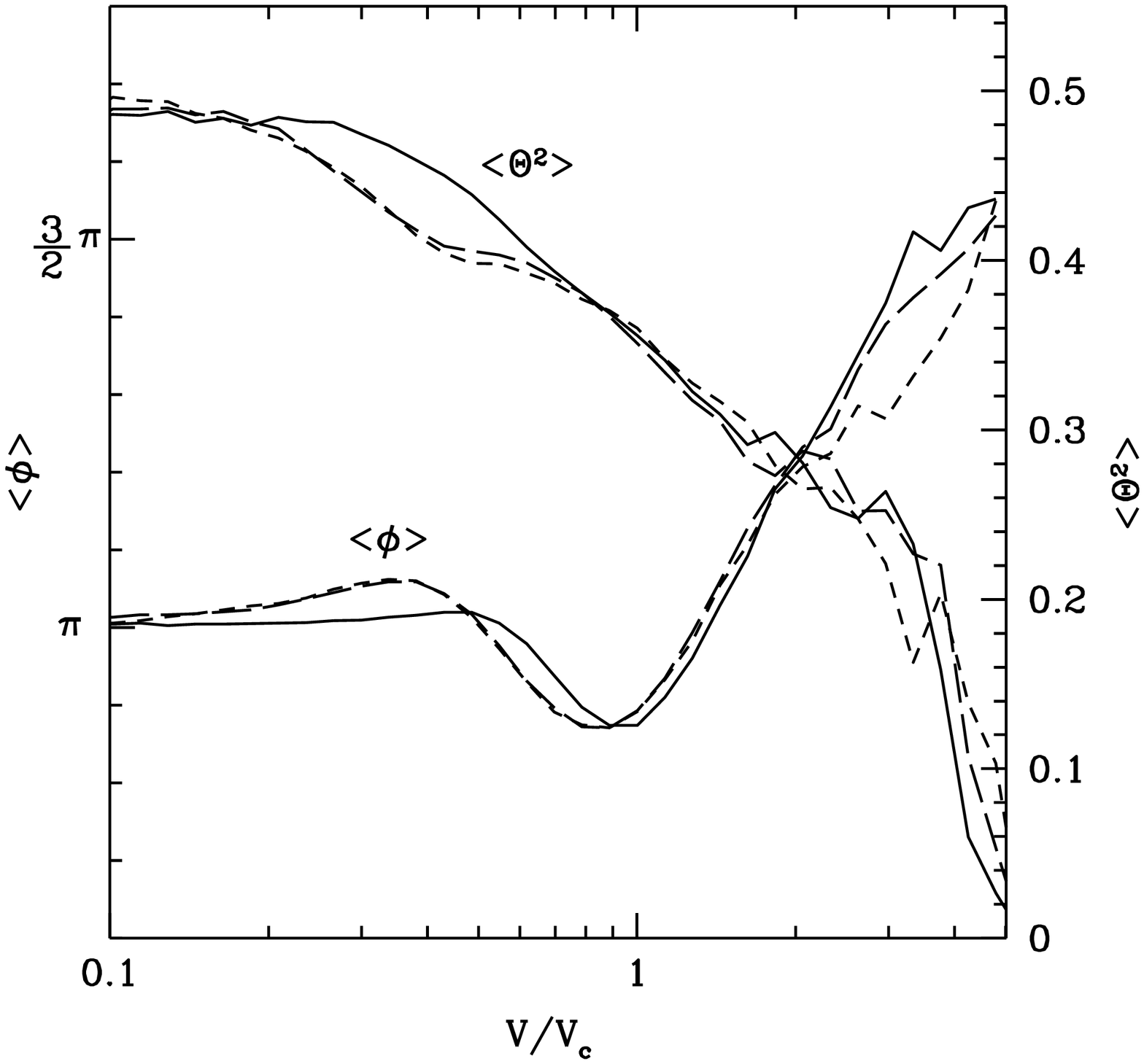,width=2.9in}}
\vspace{-0.0cm}
\caption{\footnotesize Mean values of $\theta^2$ ({\it right scale axis}) and $\phi$ ({\it 
left scale axis}) versus $V/V_c$ at different binary separations: 
$a/a_h=1$ ({\it solid lines}), $a/a_h=0.1$ ({\it long-dashed lines}), 
and $a/a_h=0.01$ ({\it short-dashed lines}). A mass ratio $q=1/9$ and an 
eccentricity $e=0.6$ are assumed.}
\label{fig:thetafivsVc}
\vspace{+0.5cm}
\end{figurehere}
\section{Conclusions}

We have performed full three-body scattering experiments in order to
study the detailed kinematic properties of hypervelocity stars by massive 
black hole binaries at the center of galaxies. Ambient stars are drawn 
from a Maxwellian distribution unbound to the binary, and are expelled by 
the gravitational slingshot. Numerical orbit integration from initial conditions 
determined by Monte Carlo techniques provides accurate measurements of thermally 
averaged hardening, mass ejection, and eccentricity growth rates for MBHBs 
in a fixed stellar background. We have shown that binary-star interactions create a 
subpopulation of HVSs on nearly radial orbits, with a spatial 
distribution 
that is initially highly flattened in the inspiral plane of the MBHB, but 
becomes more isotropic with decreasing binary separation. The degree of 
anisotropy is smaller for unequal mass binaries and larger for stars with 
higher kick velocities. Eccentric MBHBs produce a more prominent tail of 
high-velocity stars and break axisymmetry, ejecting HVSs along a 
broad jet perpendicular to the semimajor axis. The jet two-sidedness 
decreases with increasing binary mass ratio, while the jet opening- angle
increases with decreasing kick velocity and orbital separation.     

It is interesting to quantify the properties of the HVSs that would
populate the halo of the Milky-Way (MW) in the presence of a MBHB at the Galactic
center. We assume that the binary total mass is $M_1+M_2=3.5\times 10^6\,\msun$, 
that Sgr A$^*$ is the most massive component $M_1$ of the pair, that the 
binary mass ratio is $q=1/81$, and that the loss-cone is always full. From our 
discussion in \S4.1, we expect the small mass ratio to result in a large number 
of HVSs. The pair is allowed to shrink from $a=a_h$ down to $a=0.1\,a_h$ due to 
three-body interactions, within the allowed parameter space derived for a circular 
binary by Yu \& Tremaine (2003) using a variety of observational and theoretical 
arguments. Figure \ref{fig:dMdlna} shows the quantity $d(N_{\rm HVS})/d\ln(a)$, 
the number of HVSs (assuming $m_*=1\,\msun$) ejected per logarithmic binary 
separation with $V>1.68 \,v_{\rm esc}$. The adopted velocity threshold implies 
$V>840\,\kms$ at the influence radius of the binary: it corresponds to the 
escape velocity from the center of the MW galaxy,\footnote{We modeled the luminous 
component of the MW as in Miyamoto \& Nagai (1975), and the dark matter halo 
as in Widrow \& Dubinski (2005).}\ and translates within the MW potential to 
$V>450$ km/s 10 kpc away from SgrA$^*$.
\begin{figurehere}
\vspace{0.5cm} 
\centerline{\psfig{figure=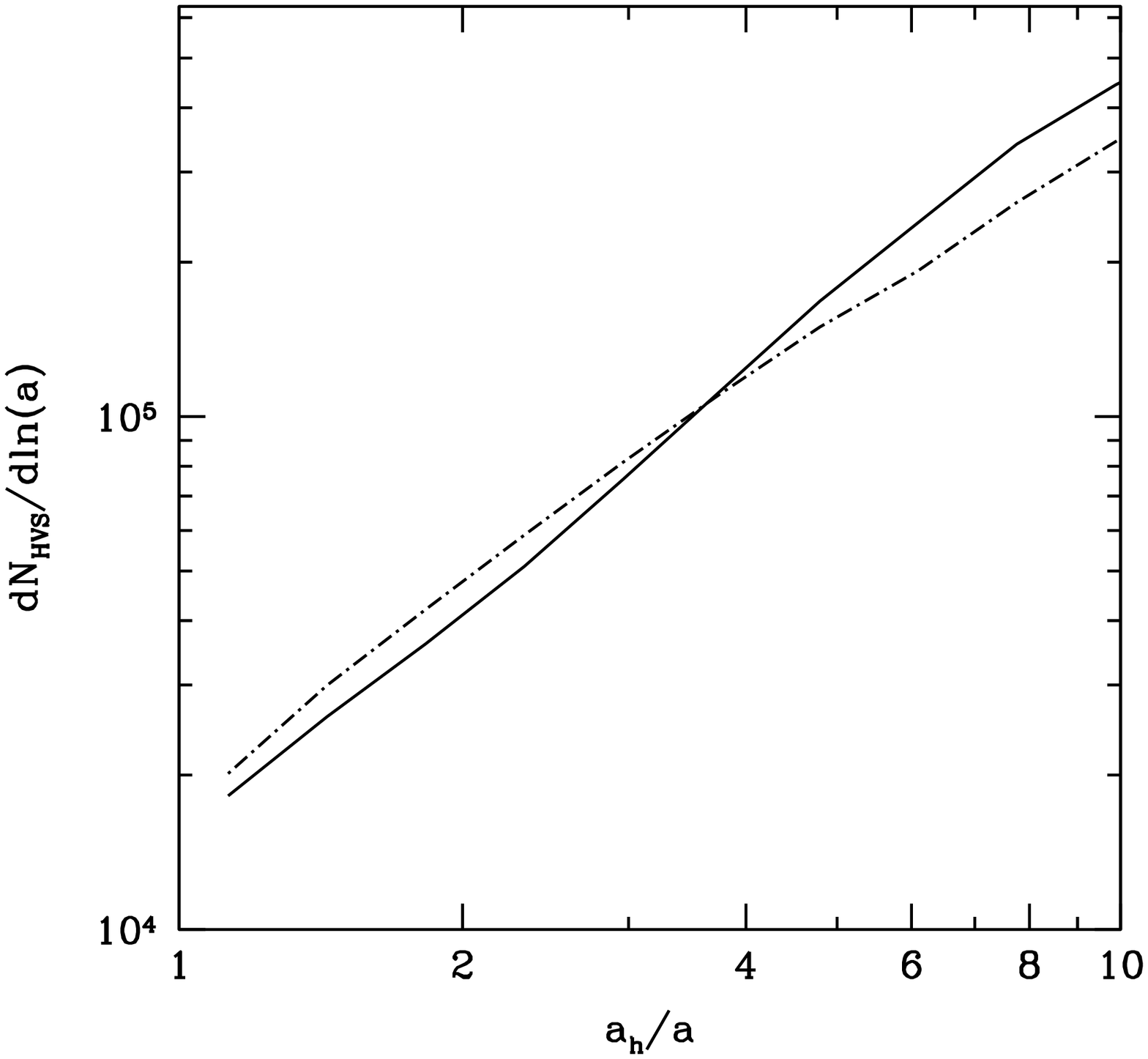,width=2.9in}}
\vspace{-0.0cm}
\caption{\footnotesize Number of HVSs escaping the MW. The binary total mass is $M_1+M_2=3.5\times 10^6\,\msun$, 
and the mass ratio is $q=M_2/M_1=1/81$. {\it Solid line:} binary with initial 
eccentricity $e=0$. {\it Dot-dashed line}: binary with initial eccentricity $e=0.9$. 
The total number of HVSs ejected as the pair shrinks from $a=a_h$ to 
$a=a_h/2$ ($a=a_h/10$) is $5\times 10^4$ ($10^6$). 
}
\label{fig:dMdlna}
\vspace{+0.5cm}
\end{figurehere}

The curves show results for different initial 
eccentricities: in a self-consistent treatment of the evolution of a MBHB, we have
used the results of our scattering experiments to account for the changing 
binary eccentricity as its orbit decays. 
As the orbital separation decreases from $a_h$ to $a_h/2$, we find that the total
number of HVSs is $5\times 10^4$, for a total ejected mass of some $1.2\,M_2$,
independent of eccentricity: in the case of a circular binary the angular distribution 
of such HVSs is characterized by $\langle \theta^2\rangle=0.3$ and 
$\sigma_\phi/\langle \phi\rangle=0.58$,
while for an eccentric pair with $e=0.9$ the latitude and azimuthal dispersions
are $\langle \theta^2\rangle=0.37$ and $\sigma_\phi/\langle \phi\rangle=0.6$.
In the case in which the binary separation shrinks from the hardening radius down 
to $a=0.1\,a_h$ instead, the total number of ejected HVSs increases by a 
factor of 20, to $10^6$: a circular binary produces an angular distribution 
with $\langle \theta^2\rangle=0.36$ and $\sigma_\phi/\langle \phi\rangle=0.58$, 
while for an eccentric pair with $e=0.9$ we find $\langle \theta^2\rangle=0.4$ 
and $\sigma_\phi/\langle \phi\rangle=0.6$.
The detection of a numerous population of HVSs in the halo of the Milky Way 
by the next generation of large astrometric surveys like GAIA may thus provide a 
clear signature of the history, nature, and environment of the MBH at the Galactic 
center.

\section*{ACKNOWLEDGMENTS}
We thank M. Colpi, M. Dotti, M. Rees, S. Tremaine, and A. Vecchio for 
discussions on the dynamics of black hole binaries. Support for this work was 
provided by NSF grant AST02-05738 and NASA grants NAG5-11513 and NNG04GK85G (P.M.), 
and by the Italian MIUR grant PRIN 2004 (A.S. and F.H.). P.M. also acknowledges 
support from the Alexander von Humboldt Foundation.

{}
\end{document}